\documentclass[doublecol]{epl2}

\title{The Visibility Graph: a new method for estimating the Hurst exponent
of fractional Brownian motion}
\shorttitle{Visibility Graph quantifies colored noise} 

\author{Lucas Lacasa\inst{1} \and Bartolo Luque\inst{1} \and Jordi Luque\inst{2} \and Juan Carlos Nu\~{n}o\inst{3}}

\institute{
  \inst{1} Dpto. de Matem\'atica Aplicada y Estad\'istica,
ETSI Aeron\'auticos,  Universidad Polit\'ecnica de Madrid,
Spain.\\
  \inst{2} Dept de Teoria del Senyal i Comunicacions, Universitat
Polit$\grave{e}$cnica de Catalunya, Spain.\\
\inst{3} Dpto. de Matem\'{a}tica Aplicada a los Recursos Naturales,
ETSI Montes, Universidad Polit\'{e}cnica de Madrid, Spain.}


\abstract{Fractional Brownian motion (fBm) has been used as a
theoretical framework to study real time series appearing in diverse
scientific fields. Because its intrinsic non-stationarity and long
range dependence, its characterization {\it via} the Hurst parameter
$H$ requires sophisticated techniques that often yield ambiguous
results. In this work we show that fBm series map into a scale free
\emph{visibility graph} whose degree distribution is a function of
$H$. Concretely, it is shown that the exponent of the power law
degree distribution depends linearly on $H$. This also applies to
fractional Gaussian noises (fGn) and generic $f^{-\beta}$ noises.
Taking advantage of these facts, we propose a brand new methodology
to quantify long range dependence in these series. Its reliability
is confirmed with extensive numerical simulations and analytical
developments. Finally, we illustrate this method quantifying the
persistent behavior of human gait dynamics.}

\begin{document}

\maketitle

Self-similar processes such as fractional Brownian motion (fBm)
\cite{fbm_mandel} are currently used to model fractal phenomena of
different nature, ranging from Physics or Biology to Economics or
Engineering. To cite a few, fBm has been used in models of
electronic delocalization \cite{fbm_condmat}, as a theoretical
framework to analyze turbulence data \cite{fbm_turb}, to describe
geologic properties \cite{fbm_geo}, to quantify correlations in DNA
base sequences \cite{fbm_dna}, to characterize physiological signals
such as human heartbeat \cite{fbm_nat} or gait dynamics
\cite{gait_review}, to model economic data \cite{fbm_eco} or to
describe network traffic \cite{IEEE-kara, fbm_traffic1,
fbm_traffic2}. Fractional Brownian motion $B_H(t)$ is a
non-stationary random process with stationary self-similar
increments (fractional Gaussian noise) that can be characterized by
the so called Hurst exponent, $0 < H <1$. The one-step memory
Brownian motion is obtained for $H= \frac{1}{2}$, whereas time
series with $H
> \frac{1}{2}$ shows persistence and anti-persistence if $H <
\frac{1}{2}$.\\

\noindent While different fBm generators and estimators have been
introduced in the last years, the community lacks consensus on which
method is best suited for each case. This drawback comes from the
fact that fBm formalism is exact in the infinite limit, i.e. when
the whole infinite series of data is considered. However, in
practice, real time series are finite. Accordingly, long range
correlations are partially broken in finite series, and local
dynamics corresponding to a particular temporal window are
overestimated. The practical simulation and the estimation from real
(finite) time series is consequently a major issue that is,
hitherto, still open. An overview of different methodologies and
comparisons can be found in \cite{IEEE-kara,physa_LRD, physd_LRD,
IEEE-Viei, SERIES_PRL, carbone, Miel,wavelet_pre} and references
therein.
 \begin{figure}[h]
 \hspace*{0.0cm}
 \includegraphics[width=0.45\textwidth]{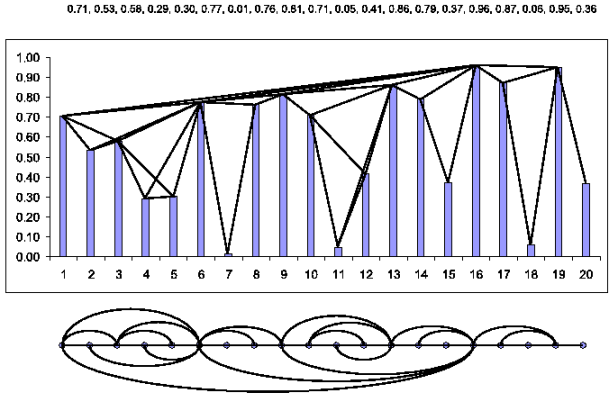}%
\caption{Example of a time series ($20$ data, depicted in the upper
part) and the associated graph derived from the visibility
algorithm. In the graph, every node corresponds, in the same order,
to a series data. The visibility rays between the data define the
links connecting nodes in the graph.}
 \label{visib}
 \end{figure}
Most of the preceding methods operate either on the time domain
(e.g. Aggregate Variance Method, Higuchi's Method, Detrended
Fluctuation Analysis, Range Scaled Analysis, etc) or in the
frequency or wavelet domain (Periodogram Method, Whittle Estimator,
Wavelet Method). In this letter we introduce an alternative and
radically different method, the Visibility Algorithm, based in graph
theoretical techniques. In a recent paper this new tool for
analyzing time series has been presented \cite{PNAS}. In short, a
visibility graph is obtained from the mapping of a time series into
a network according with the following visibility criterium: two
arbitrary data $(t_a,y_a)$ and $(t_b,y_b)$ in the time series have
visibility, and consequently become two connected nodes in the
associated graph, if any other data $(t_c,y_c)$ such that $t_a < t_c
< t_b$ fulfills:
\begin{equation}
y_c < y_b + (y_a - y_b) \frac{t_b - t_c}{t_b - t_a}. \label{eq1}
\end{equation}
In fig.\ref{visib} we have represented for illustrative purposes an
example of how a given time series maps into a visibility graph by
means of the Visibility Algorithm. A preliminary analysis has shown
that series structure is inherited in the visibility graph
\cite{PNAS}. Accordingly, periodic series map into regular graphs,
random series into random graphs and fractal series into scale free
graphs \cite{barabasi}. In particular, it was shown that the
visibility graph obtained from the well-known Brownian motion has
got both the scale-free and the small world properties \cite{PNAS}.
Here we show that the visibility graphs derived from generic fBm
series are also scale free. This robustness goes further, and we
prove that a linear relation between the exponent $\gamma$ of the
power law degree distribution in the visibility graph and the Hurst
exponent $H$ of the associated fBm series exists. Therefore, the
visibility algorithm provides an alternative method to compute the
Hurst exponent and then, to characterize fBm processes. This also
applies to fractional gaussian noise (fGn) \cite{fbm_mandel} which
are nothing but the increments of a fBm, and generic $f^{-\beta}$
noises, enhancing the visibility graph as a
method to detect long range dependence in time series.\\
\begin{figure}[h]
\includegraphics[width=0.45\textwidth]{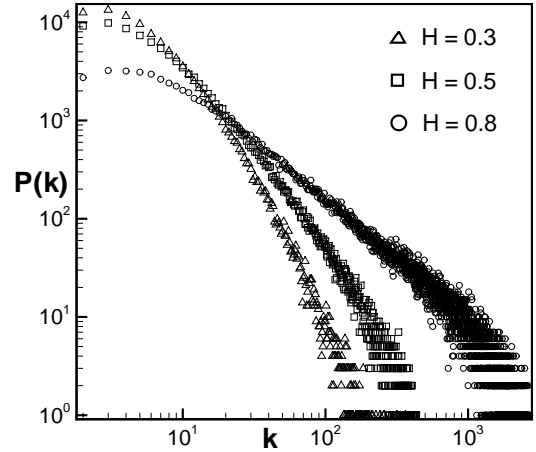} \caption{Degree
distribution of three visibility graphs, namely (i) triangles:
extracted from a fBm series of $10^5$ data with $H=0.3$, (ii)
squares: extracted from a fBm series of $10^5$ data with $H=0.5$,
(iii) circles: extracted from a fBm series of $10^5$ data with
$H=0.8$. Note that distributions are not normalized. The three
visibility graphs are scale-free since their degree distributions
follow a power law $P(k)\sim k^{-\gamma}$ with decreasing exponents
$\gamma_{0.3}>\gamma_{0.5}>\gamma_{0.8}$.} \label{figure1}
\end{figure}
In fig.\ref{figure1} we have depicted in log-log the degree
distribution of the visibility graph associated with three
artificial fBm series of $10^5$ data, namely an anti-persistent
series with $H=0.3$ (triangles),
 a memoryless Brownian motion with $H=0.5$ (squares) and a persistent fBm with $H=0.8$ (circles). As
can be seen, these distributions follow a power law $P(k) \sim k^{-
\gamma}$ with decreasing exponents
$\gamma_{0.3}>\gamma_{0.5}>\gamma_{0.8}$.
\begin{figure}[h]
\includegraphics[width=0.45\textwidth]{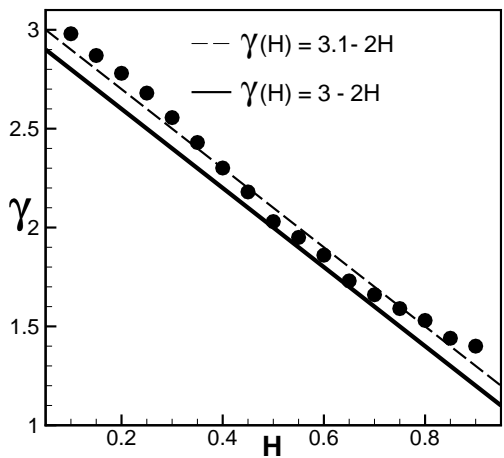} \caption{(Black
dots) Numerical estimation of exponent $\gamma$ of the visibility
graph associated with a fBm series with exponent $H$. In each case
$\gamma$ is averaged over $10$ realizations of a fBm series of
$10^4$ data, in order to avoid non-stationary biases (the error bars
are included in the dot size). The dotted line corresponds to the
best linear fitting $\gamma(H)=a-bH$, where $a=3.1\pm0.1$ and
$b=2.0\pm0.1$, and the solid line corresponds to the theoretical
prediction $\gamma(H)=3-2H$. Both results are consistent. Note that
deviations from the theoretical law take place for values of $H>0.5$
and $H<0.5$ (strongly correlated or anti-correlated series), where
fBm generators evidence finite-size accuracy problems \cite{Horn},
these being more acute the more we move away from the non-correlated
case $H=0.5$.} \label{figure2}
\end{figure}

\noindent In order to compare $\gamma$ and $H$ appropriately, we
have calculated the exponent of different scale free visibility
graphs associated with fBm artificial series of $10^4$ data with
$0<H<1$ generated by a wavelet based algorithm \cite{matlab}. Note
at this point that some bias is inevitably present since artificial
series generators are obviously not exact, and consequently the
nominal Hurst exponents have an associated error \cite{Horn}. For
each value of the Hurst parameter we have thus averaged the results
over $10$ realizations of the fBm process. We have estimated
exponent $\gamma$ in each case through Maximum Likelihood Estimation
(MLE) \cite{newmann}:
\begin{equation}
\gamma=1+n\bigg[\sum_{i=1}^n\log\frac{x_i}{x_{min}}\bigg]^{-1},
 \label{mle}
\end{equation}
where $n$ is total number of values taken into account, $x_i,\
i=1,..,n$ are the measured values and $x_{min}$ corresponds to the
smallest value of $x$ for which the power law behavior holds.
 In fig.\ref{figure2} we have represented the
relation between $\gamma$ and $H$ (black circles). As can be seen, a
roughly linear relation holds (the dotted line represents the best
linear fitting
$\gamma=3.1-2H$).\\
\begin{figure}[h] \includegraphics[width=0.45\textwidth]{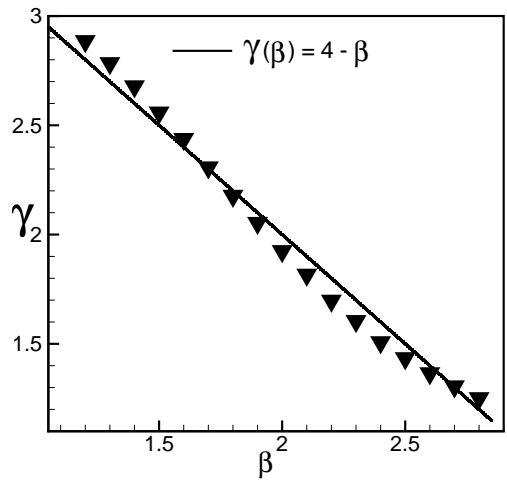} \caption{(Black triangles) Numerical
estimation of exponent $\gamma$ of the visibility graph associated
with a $f^{-\beta}$ noise. In each case $\gamma$ is averaged over
$10$ realizations of a $f^{-\beta}$ series of $10^6$ data, in order
to avoid non-stationary biases (the error bars are included in the
triangle size). The straight line corresponds to the theoretical
prediction in eq.\ref{teorico3}.} \label{figure3}
\end{figure}
\noindent That fBm yields scale free visibility graphs is not that
surprising. The most highly connected nodes (hubs) are the
responsible for the heavy tailed degree distributions. Within fBm
series, hubs are related to extreme values in the series, since a
data with a very large value has typically a large connectivity,
according to eq. \ref{eq1}. In order to calculate the tail of the
distribution we consequently need to focus on the hubs, and thus
calculate the probability that an extreme value has a degree $k$.
Suppose that at time $t$ the series reaches an extreme value (a hub)
$B_H(t)=h$. The probability of this hub to have degree $T$ is
\begin{equation}
P(T)\sim P_{fr}(T)r(T),\label{heuristic}
\end{equation}
where $P_{fr}(T)$ provides the probability that after $T$ time
steps, the series returns to the same extreme value, i.e. $B(t+T)=h$
(and consequently the visibility in $t$ gets truncated in $t+T$),
and $r(T)$ is the percentage of nodes between $t$ and $t+T$ that $t$
may see. $P_{fr}(T)$ is nothing but the first return time
distribution, which is known to scale as $P_{fr}(T)\sim T^{H-2}$ for
fBm series \cite{PRE_ding}. On the other hand, the percentage of
visible nodes between two extreme values is related to the roughness
of the series in that basin, that is, to the way that a series of
$T$ time steps folds. This roughness is encoded in the series
standard deviation \cite{fbm_mandel}, such that intuitively, we have
$r(T)\sim T^H/T=T^{H-1}$ (this fact has been confirmed numerically).
Finally, notice that in this context $T\equiv k$, so
eq.\ref{heuristic} converts into
\begin{equation}
P(k)\sim k^{H-2}k^{H-1}=k^{2H-3}, \label{teorico}
\end{equation}
what provides a linear relation between the exponent of the
visibility graph degree distribution and the Hurst exponent of the
associated fBm series:
\begin{equation}
\gamma(H)=3-2H, \label{teorico2}
\end{equation}
in good agreement with our previous numerical results. Note in
figure \ref{figure2} that numerical results obtained from artificial
series deviate from the theoretical prediction for
strongly-correlated ones ($H>0.5$ or $H<0.5$). This deviation is
related to finite size effects in the generation of finite fBm
series \cite{Horn}, and these effects are more acute the more we
deviate from the non-correlated case $H=0.5$. In any case, a scatter
plot of the theoretical (eq.\ref{teorico2}) versus the empirical
estimation of $\gamma(H)$ provides
statistical conformance with a correlation coefficient $c=0.99$.\\
\noindent To check further the consistency of the visibility
algorithm, an estimation of the power spectra is performed. It is
well known that fBm has a power spectra that behaves as
$1/f^{\beta}$, where the exponent $\beta$ is related to the Hurst
exponent of an fBm process through the well known relation
\cite{Adison}
\begin{equation}
\beta(H)=1+2H. \label{beta}
\end{equation}
Now according to eqs.\ref{teorico2} and \ref{beta}, the degree
distribution of the visibility graph corresponding to a time series
with $f^{-\beta}$ noise should be again power law $P(k)\sim
k^{-\gamma}$ where
\begin{equation}
\gamma(\beta)=4-\beta. \label{teorico3}
\end{equation}
In fig.\ref{figure3} we depict (triangles) the empirical values of
$\gamma$ corresponding to $f^{-\beta}$ artificial series of $10^6$
data with $\beta$ ranging from $1.2$ to $2.8$ in steps of size $0.1$
\cite{ruidof}. For each value of $\beta$ we have again averaged the
results over $10$ realizations and estimated $\beta$ through MLE
(eq.\ref{mle}). The straight line corresponds to the theoretical
prediction eq.\ref{teorico3}, showing good agreement with the
numerics. In this case, a scatter plot confronting theoretical
versus empirical estimation of $\gamma(\beta)$ also provides
statistical conformance between them, up to $c=0.99$.\\
Finally, observe that eq.\ref{beta} holds for fBm processes, while
for the increments of an fBm process, known as a fractional Gaussian
noise (fGn), the relation between $\beta$ and $H$ turns to be
\cite{Adison}:
\begin{equation}
\beta(H)=-1+2H, \label{beta2}
\end{equation}
where $H$ is the Hurst exponent of the associated fBm process. We
consequently can deduce that the relation between $\gamma$ and $H$
for a fGn (where fGn is a series composed by the increments of a
fBm) is
\begin{equation}
\gamma(H)=5-2H. \label{gamma2}
\end{equation}
Notice that eq.\ref{gamma2} can also be deduced applying the same
heuristic arguments as for eq.\ref{teorico2} with the change
$H\rightarrow H-1$.\\
\noindent In order to illustrate this latter case, we finally
address a realistic and striking dynamics where long range
dependence has been recently described. Gait cycle (the stride
interval in human walking rhythm) is a physiological signal that has
been shown to display fractal dynamics and long range correlations
in healthy young adults \cite{pnas_dfa, gait}. In the upper part of
fig.\ref{figure4} we have plotted to series describing the
fluctuations of walk rhythm of a young healthy person, for slow pace
(bottom series of $3304$ points) and fast pace (up series of $3595$
points) respectively (data available in
www.physionet.org/physiobank/database/umwdb/ \cite{physionet}). In
the bottom part we have represented the degree distribution of their
visibility graphs. These ones are again power laws with exponents
$\gamma=3.03\pm0.05$ for fast pace and $\gamma=3.19\pm0.05$ for slow
pace (derived through MLE). According to eq.\ref{teorico3}, the
visibility algorithm predicts that gait dynamics evidence
$f^{-\beta}$ behavior with $\beta=1$ for fast pace, and $\beta=0.8$
for slow pace, in perfect agreement with previous results based on a
Detrended Fluctuation Analysis \cite{pnas_dfa, gait}. These series
record the fluctuations of walk rhythm (that is, the increments), so
according to eq.\ref{gamma2}, the Hurst exponent is $H=1$ for fast
pace and $H=0.9$ for slow pace,
that is to say, dynamics evidences long range dependence (persistence) \cite{pnas_dfa, gait}.\\
\begin{figure}[h]
\leavevmode \includegraphics[width=0.45\textwidth]{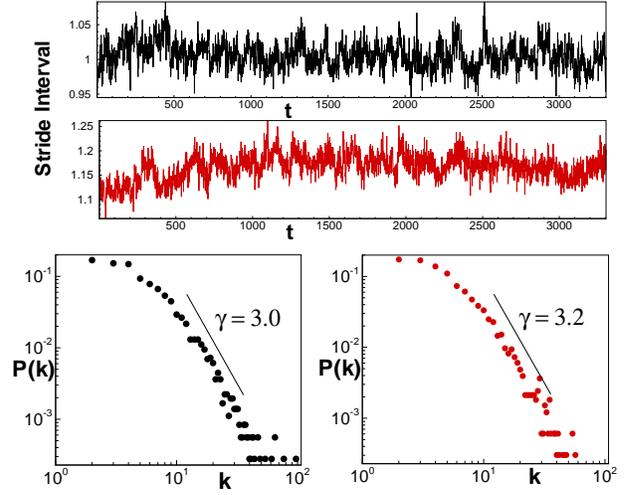}
\caption{Black signal: time series of $3595$ points from the stride
interval of a healthy person in fast pace. Red signal: time series
of $3304$ points from the stride interval of a healthy person in
slow pace. Bottom: Degree distribution of the associated visibility
graphs (the plot is in log-log). These are power laws where
$\gamma=3.03\pm0.05$ for the fast movement (black dots) and
$\gamma=3.19\pm0.05$ for the slow movement (red dots), what provides
$\beta=1$ and $\beta=0.8$ for fast and slow pace respectively
according to eq.\ref{teorico3}, in agreement with previous results
\cite{pnas_dfa, gait}.} \label{figure4}
\end{figure}

\noindent As a summary, the visibility graph is an algorithm that
map a time series into a graph. In so doing, classic methods of
complex network analysis can be applied to characterize time series
from a brand new viewpoint \cite{PNAS}. In this work we have pointed
out how graph theory techniques can provide an alternative method to
quantify long range dependence and fractality in time series. We
have reported analytical and numerical evidences showing that the
visibility graph associated to a generic fractal series with Hurst
exponent $H$ is a scale free graph, whose degree distribution
follows a power law $P(k)\sim k^{-\gamma}$ such that: (i) There is a
universal relation between $\gamma$ and the exponent $\beta$ of its
power spectrum that reads $\gamma=4-\beta$; (ii) for fBm signals
(where $H$ is defined such that $\beta(H)=1+2H$), the relation
between $\gamma$ and H reads $H(\gamma)=\frac{3-\gamma}{2}$ while
for fGn signals (the increments
of a fBm where H is defined as $\beta(H)=-1+2H$), we have $H(\gamma)=\frac{5-\gamma}{2}$.\\
The reliability of this methodology has been confirmed with
extensive simulations of artificial fractal series and real (small)
series concerning gait dynamics. To our knowledge, this is the first
method for estimation of long range dependence in time series based
in graph theoretical techniques advanced so far. Some questions
concerning its accuracy, flexibility and computational efficiency
will be at the core of further investigations. In any case, we do
not pretend in this work to compare its accuracy with other
estimators, but to propose an alternative and simple method based in
completely different techniques with potentially broad applications.

\acknowledgments
The authors thank Octavio Miramontes and Fernando
Ballesteros for helpful suggestions. This work was partially
supported by Spanish Ministry of Science Grant FIS2006-08607.

\end{document}